\documentclass{article}

\PassOptionsToPackage{square, comma, numbers, sort}{natbib}

\usepackage[preprint]{neurips_2024}

\usepackage[utf8]{inputenc} 
\usepackage[T1]{fontenc}    
\usepackage{hyperref}       
\usepackage{url}            
\usepackage{booktabs}       
\usepackage{amsfonts}       
\usepackage{nicefrac}       
\usepackage{microtype}      
\usepackage{xcolor}         
\usepackage[ruled]{algorithm2e}
\usepackage{color,url,subfigure,graphicx,wrapfig}
\usepackage{amsmath}
\usepackage{enumerate}
\usepackage{enumitem}
\usepackage{verbatim}
\usepackage{footnote}
\newtheorem{theorem}{Theorem}

\newtheorem{assumption}{Assumption}

\newtheorem{proof}{Proof}[section]

\title{Understanding Byzantine Robustness in Federated Learning with A Black-box Server}
\renewcommand{\thefootnote}{\fnsymbol{footnote}}

\author{Fangyuan Zhao$^{1,\star}$, Yuexiang Xie$^2$, Xuebin Ren$^1$, Bolin Ding$^2$, Shusen Yang$^{1,\dagger}$, Yaliang Li$^{2,\dagger}$\\$^1$Xi'an Jiaotong University\ \ $^2$Alibaba Group}

\begin{document}

\footnotetext[1]{Work done at Alibaba Group.}
\footnotetext[2]{Corresponding authors.}

\maketitle
\renewcommand{\thefootnote}{\arabic{footnote}}

\begin{abstract}
Federated learning (FL) becomes vulnerable to Byzantine attacks where some of participators tend to damage the utility or discourage the convergence of the learned model via sending their malicious model updates. Previous works propose to apply robust rules to aggregate updates from participators against different types of Byzantine attacks, while at the same time, attackers can further design advanced Byzantine attack algorithms targeting specific aggregation rule when it is known. In practice, FL systems can involve a black-box server that makes the adopted aggregation rule inaccessible to participants, which can naturally defend or weaken some Byzantine attacks. In this paper, we provide an in-depth understanding on the Byzantine robustness of the FL system with a black-box server. Our investigation demonstrates the improved Byzantine robustness of a black-box server employing a dynamic defense strategy. We provide both empirical evidence and theoretical analysis to reveal that the black-box server can mitigate the worst-case attack impact from a maximum level to an expectation level, which is attributed to the inherent inaccessibility and randomness offered by a black-box server. 
The source code is available\footnote{\url{https://github.com/alibaba/FederatedScope/tree/Byzantine\_attack\_defense}} to promote further research in the community.
\end{abstract}

\section{Introduction}
\label{sec:intro}

Federated learning (FL)~\cite{kairouz2021advances, bonawitz2019towards, ren2024belt} enables massive clients to collaboratively train a global model without private data exposure, providing a solution to improving the model utility by sharing learned knowledge in a privacy-preserving manner. In an FL system, a server is expected to periodically aggregate the model updates provided by multiple clients, and then to optimize the global model accordingly. The most widely-used FL algorithm \texttt{Fedavg}~\cite{li2019convergence} takes the \texttt{mean} operator in federated aggregation. However, it has been pointed out by previous studies~\cite{baruch2019little,fang2020local,shejwalkar2021manipulating} that the global model learned by \texttt{Fedavg} can be vulnerable when some Byzantine clients tend to send their malicious model updates. 
The intrinsic reason is that \texttt{Fedavg} equally takes model updates provided by all the clients, including the malicious ones, but the \texttt{mean} operation adopted in federated aggregation might not be robust enough against malicious updates.

Recent studies~\cite{blanchard2017machine,yin2018byzantine,cao2020fltrust} propose to apply robust aggregation rules for enhancing the Byzantine robustness of FL systems.
For example, Krum~\cite{blanchard2017machine} suggests filtering outliers in the model updates based on Euclidean distances; Median~\cite{yin2018byzantine} adopts the statistics \texttt{median} operator rather than the \texttt{mean} operator in federated aggregation. These proposed robust rules make remarkable progress in effectively defending some Byzantine attacks that malicious updates can be generated without additional knowledge of aggregation rules adopted at the server, which are called \texttt{AGR-agnostic} attacks in this study.

Meanwhile, some advanced attack algorithms~\cite{fang2020local,shejwalkar2021manipulating,shejwalkar2022back} are designed targeting specific aggregation rules, which might easily disable the defense and severely damage the utility of the learned global model when the aggregation rules are known. These Byzantine attacks, named \texttt{AGR-adaptive} attacks, aim to optimize malicious updates that can both cheat the adopted robust aggregation rules and incur significant negative impact on the model performance. 
For example, {\it Fang attack}~\cite{fang2020local} scales the inverse direction of the model updates to generate malicious ones for maximizing the inner product between the aggregate deviation and the perturbation direction. 

In real-world applications, it is reasonable for FL systems to involve a black-box server that makes the adopted robust aggregation rules inaccessible to attackers, which can violate the assumption that \texttt{AGR-adaptive} attacks rely on, i.e., the knowledge of the aggregation rule is accessible. 
It fits an intuition that using a black-box server can naturally defend or weaken some Byzantine attacks, but there still remains a gap between intuition and an in-depth understanding of the robustness of the FL system with a black-box server.

To fulfill such a gap, we conduct the first study on Byzantine robustness of the FL system with a black-box server, providing both theoretical analysis and empirical observations for in-depth understanding. 
Specifically, we conduct the study to answer the question: \textit{Compared to white-box server settings (i.e., the behaviors are exposed to malicious clients), how and to what extent a black-box server can enhance the Byzantine robustness of an FL system?}

We set up an \texttt{attack and defense} model to compare the white-box server and the black-box server scenarios, applying static and dynamic defense strategies, respectively. 
With the knowledge of the robust aggregation rules, Byzantine attacks on the white-box server can always incur the worst-case impact when malicious clients adopt the targeting Byzantine attack algorithms.
We point out that such vulnerability mainly comes from unequal positions in the game of attack and defense under the white-box server setting. In contrast, a black-box server can mitigate such vulnerability as the black-box property takes the server to an equal position with the hidden Byzantine clients.

We provide theoretical analysis on robustness and convergence for dynamic defense strategies. Accordingly, we further analyze the improvement of Byzantine robustness when employing a black-box server as opposed to a white-box server, showing that a black-box server can reduce the worst-case attack impact from a maximum level to an expectation level. 
The reduction essentially sources from the fact that a white-box server consistently provides prior knowledge to malicious clients, enabling them to find an optimal attack strategy, whereas a black-box server can effectively prevent the exposure of any useful information about the defense strategy.
We conduct a series of experiments on widely-used datasets, showing that a black-box server can significantly enhance the robustness of the FL systems against various Byzantine attacks by adopting the dynamic defense strategy.

\section{Preliminary}
\label{sec:preliminary}
In this section, we present some preliminary information on federated learning and Byzantine robustness, and formulate the considered Byzantine attacks in this study.

\paragraph{Federated Learning}
Let $\mathcal{F}(x;D)$ be the objective function to be minimized where $D$ is the underlying data distribution and $x\in \mathbb{R}^d$ represents the model parameters. For simplicity, we use $\mathcal{F}(x)$ to denote $\mathcal{F}(x;D)$ subsequently. $f(x;\xi)$ is an approximation to $\mathcal{F}(x)$ under a federated learning setting in which $\xi\sim D$ is an observed data distributed among the clients in an FL course. Each client $c_i$ owns a part of dataset $\xi^i$ and optimizes an objective function $f_i(x;\xi_i)$ (also denoted as $f_i(x)$ subsequently). The federated learning task can be formulated as an optimization problem as:
\begin{equation}\label{equ:objective_func}
    \min_{x}\mathcal{F}(x)\approx f(x;\xi)= \frac{1}{n}\sum_{i=1}^{n}f_i(x;\xi_i).
\end{equation}

\texttt{Fedavg}~\cite{mcmahan2017communication} (Algorithm~\ref{alg:fedavg} in Appendix~\ref{sec:fedavg}) is the standard FL algorithm for solving the optimization problem defined in Equation~\eqref{equ:objective_func}, which takes the weighted {\tt mean} operation over all the local updates from clients to approximate the real model update of $\mathcal{F}(x)$ on $D$.

\paragraph{Byzantine Robustness}

We follow the definition of Byzantine robustness in previous studies~\cite{blanchard2017machine,allouah2023fixing, farhadkhani2022byzantine}. Specifically, given an integer value $0< h< n/2$ and a real value $\alpha\geq 0$, assume that $V_1, ..., V_n$ are $n$ random vectors, an aggregation rule $\mathcal{AGR}$ is said to be $(h, \alpha)$-robust if for any $1\leq j_1<...<j_{h}\leq n$, vector
\begin{equation*}
    Q = \mathcal{AGR}(V_1,..., \underbrace{B_1}_{j_1}, ..., \underbrace{B_h}_{j_h}, ..., V_n)
\end{equation*}
satisfies that for any set $\mathcal{N}\subset [n]$ of size $n-h$, 
(i) $\langle Q, \bar{V}_{\mathcal{N}}\rangle > 0$; and 
(ii)
$    \|Q - \bar{V}_{\mathcal{N}}\|^2 \leq \frac{\alpha}{|\mathcal{N}|}\sum_{i\in \mathcal{N}}\|V_i-\bar{V}_{\mathcal{N}}\|^2
$
where $\bar{V}_{\mathcal{N}} = \frac{1}{|\mathcal{N}|}\sum_{i\in \mathcal{N}}V_i$ and $[n] = \{1, \ldots, n\}$.

In the above definition, condition (i) ensures that the aggregated result remains an acute angle with the averaged results, while condition (ii) quantifies the Byzantine robustness with two factors: a tolerant Byzantine client number $h$ and a robustness coefficient $\alpha$. Therefore, $(h, \alpha)$-Byzantine robustness ensures that the distance between the result of the aggregation rule $\mathcal{AGR}$ and the average of the honest inputs is bounded by $\alpha$ times the variance of all honest inputs.

\paragraph{Byzantine Attacks in FL}
Consider an FL system consisting of one central server and $n$ local clients indexed by $\{1, ..., n\}$, among which $h$ clients (indexed by $\{1, ..., h\}$) are malicious clients who can communicate with each other (i.e., collusive), and the rest $n-h$ clients are benign workers who can only communicate with the central server. In each round of training, the server first selects a set of clients to distribute the up-to-date global model. Each selected client $c_i$ then conducts the local model training on their own dataset $\xi_i$ and obtains the model updates $\Delta x^t_i$. After the local training process, each client $c_i$ uploads a vector $v$ to the server with the following rule: 
\begin{equation}\label{equ:Byzantine_update}
v=\left\{
\begin{aligned}
\Delta x^t_i & , & i>h; \\
b_i & , & i\leq h,
\end{aligned}
\right.
\end{equation}
where $b_i$ denotes a malicious model update that could be generated by colluding with all malicious clients. Under such cases, the \texttt{mean} operations adopted in \texttt{Fedavg} may not well approximate the real model update of $\mathcal{F}(x)$ on $D$.

\paragraph{Assumptions}
We give some basic assumptions~\cite{liu2023byzantine,karimireddy2021learning,karimireddy2021byzantine} in theoretical analysis.

\begin{assumption}[Unbiased stochastic gradient]\label{assum:unbiased}
    The stochastic gradient $\nabla f_i(x; \xi^{t}_{i})$ is an unbiased estimator of the local gradient $\nabla f_{i}(x)$, where $\xi^{t}_{i}$ is the sampled data of client $c_i$ at the $t$-th round,
    \begin{equation}
      \mathbb{E}_{\xi^{t}_{i}} \left[ \nabla f_i(x; \xi^{t}_{i}) \right] = \nabla f_{i} (x).
    \end{equation}
\end{assumption}

\begin{assumption}[Bounded variance]\label{assum:bound_var}
    The variance of stochastic gradients is bounded: $\exists G_{l}^{2} \in \mathbb{R}$,
    \begin{equation}
    s.t., \quad \mathbb{E}_{\xi^{t}_{i}}  [\|\nabla f_i(x, \xi^{t}_{i}) - \nabla f_i(x)\|^2]  \leq G_{l}^{2}.
    \end{equation}
\end{assumption}

\begin{assumption}[Bounded data heterogeneity]\label{assum:bound_data_hetero}
    The difference between $\nabla f_i(x)$ and $\nabla \mathcal{F}(x)$ is uniformly bounded for all benign clients: $\exists G_{g}^{2} \in \mathbb{R}$,
    \begin{equation}
     s.t., \quad \|\nabla f_i(x) - \nabla \mathcal{F}(x)\|^2 \leq G_{g}^{2}, \quad \forall i\in[n], x\in \mathbb{R}^d.
    \end{equation}
\end{assumption}

\begin{assumption}[L-smoothness]\label{assum:L-smoothness}
$\mathcal{F}(x)$ is differentiable and $L$-smooth:
\begin{equation}
    \mathcal{F}(x)-\mathcal{F}(x')\leq \langle\nabla \mathcal{F}(x), x-x'\rangle + \frac{L}{2}\Vert x-x'\Vert^2.
\end{equation}
\end{assumption}

\section{Byzantine Robustness in FL with A Black-box Server}\label{sec:Robustness within Black-box Server}

\subsection{{\tt Attack and Defense} Model}

With the aim of building an understanding of Byzantine robustness of FL systems with a black-box server, we first set up an \texttt{attack and defense} model.
In an FL system, the central server can choose to apply robust aggregation rules from the candidate set $\mathcal{S} = \{\mathcal{AGR}_i, i\in [M]\}$, where $M$ denotes the total number of candidate aggregation rules. Correspondingly, for each $\mathcal{AGR}_i\in\mathcal{S}$, malicious clients can perform \texttt{AGR-adaptive} attack $\mathcal{A}_i$ targeting $\mathcal{AGR}_i$. 
The attack $\mathcal{A}_i$ achieves the highest attack impact on the performance of $\mathcal{AGR}_i$ compared to all other attacks.

Following Byzantine robustness defined in Section~\ref{sec:preliminary}, the robustness of $\mathcal{AGR}_j$ against $\mathcal{A}_i$ can be denoted by $(h, \alpha_{i,j})$. For convenience, we fix the tolerant Byzantine client number $h$ and simply use the robust coefficient $\alpha_{i,j}$ to represent the robustness. $\mathcal{A}_i$ has the highest attack impacts on the performance of $\mathcal{AGR}_i$, thus it holds that $\alpha_{i,i} \geq \alpha_{j,i}, \forall j\neq i$.

\subsection{Static Defense Strategies}\label{subsec:fix}

For a white-box server whose aggregation rule is accessible to malicious clients, it is straightforward that whatever robust aggregation rule $\mathcal{AGR}_i$ is chosen by the white-box server, malicious clients can easily perform the attack algorithm $\mathcal{A}_i$ accordingly. As a result, the attacks incur the robust level of $\alpha_{i,i}$, which severely hurt the model performance. The vulnerability of the white-box server is essentially from its unequal position compared to hidden malicious clients in the game of defense and attack, in which any defense strategy of the white-box server is known by malicious clients in prior.

For a black-box server whose aggregation rule is inaccessible to malicious clients, the straight-forward strategy for malicious clients is to choose an arbitrary attack algorithm from $\mathcal{A}_j\in\mathcal{A}$, and verify which one can incur the maximal negative attack impact.
In such cases, whatever the strategy taken by malicious clients, the black-box server can perform no worse than the white-box server in terms of the effect of defense, because \textit{the black-box property takes the server to an equal position with hidden malicious clients in the game of defense and attack.}

In a nutshell, FL systems with a black-box server can naturally weaken Byzantine attacks compared to those with a white-box server. However, although the aggregation rule of the black-box server, say $\mathcal{AGR}_i$, is inaccessible, it is still possible that {\it the malicious clients successfully find the optimal attack method $\mathcal{A}_i$ by conducting multiple trials with different attack algorithms.}
As a result, using a static defense strategy might cause a risk that the aggregation rules are identified by the malicious clients. Therefore, black-box servers have better use a dynamic strategy to make the defense more unpredictable, which is introduced in detail below.

\subsection{Dynamic Defense Strategies}\label{subsec:dds}

Motivated by the aforementioned insights, we further consider dynamic defense, which combines diverse robust aggregation rules and incorporates randomness into the aggregation procedure to achieve unpredictability of the defense strategy in an FL system. A direct choice to introduce randomness is sampling from a set of robust aggregation rules. Specifically, the central server maintains a candidate set of robust aggregation rules $\mathcal{S} = \{\mathcal{AGR}_i, i\in [M]\}$. At each training round, the server samples a robust aggregation rule $\mathcal{AGR}_i$ from the candidate set $\mathcal{S}$ to perform aggregation according to a specific probability distribution $P = [p_1,\ldots, p_M]$. 
Both benign and malicious clients conduct the local training process and upload their updates following the definition in Equation~\eqref{equ:Byzantine_update}. 

The above sampling based dynamic defense strategy can easily achieve unpredictability, which benefits from (i) the randomness of the aggregation rule adopted in each training round, and (ii) the diversity and inaccessibility of the candidate set of aggregation rules for black-box server settings.  Therefore, this dynamic defense strategy can serve as a tool to compare the robustness of white-box and black-box server settings. Here, we first build a theoretical foundation for this dynamic defense strategy in terms of both robust level and convergence performance (the detailed proof can be found in the Appendix \ref{sec:proof}).

\begin{theorem}[Robustness Analysis]\label{theorem:robust}
Let $B = \{B_1, \ldots, B_h\}$ be an attack which can successfully attack a subset $\mathcal{S}^{\prime}$ of $q$ aggregation rules in $\mathcal{S}$, W.L.O.G., $\mathcal{S}^{\prime} = \{\mathcal{AGR}_j, j\in[q]\}$, in the sense that $\forall \mathcal{AGR}_j \in \mathcal{S}^{\prime}$, it holds that $\langle\mathcal{AGR}_j(V, B), \bar{V}\rangle \leq 0$, then the dynamic defense strategy is Byzantine robust if the probability mass on other $M-q$ aggregation rules satisfies that
\begin{equation}
\small
\begin{aligned}
\label{equ:pro_mass}
    \sum_{i\notin [q]}p_i > \frac{\sup_{j\in [q]}{|\langle Q_j, \bar{V}\rangle|}}{\sup_{j\in [q]}{|\langle Q_j, \bar{V}\rangle|} + \inf_{i\notin [q]}{\langle Q_i, \bar{V}\rangle}}
\end{aligned}
\end{equation}
where $Q_i = \mathcal{AGR}_i(V, B)$, and the robust level is $(h, \mathbb{E}_{\mathcal{AGR}_{i}\sim p}[\alpha_i])$ in expectation.
\end{theorem}

Theorem~\ref{theorem:robust} provides a sufficient condition on the candidate set $\mathcal{S}$ for achieving Byzantine robustness. The probability mass on the aggregation rules robust to the attack should be larger than a threshold which can be reduced by a larger $\inf_{i\notin [q]}{\langle Q_i, \bar{V}\rangle}$ and a smaller $\sup_{j\in [q]}{|\langle Q_j, \bar{V}\rangle|}$.  Here $\inf_{i\notin [q]}{\langle Q_i, \bar{V}\rangle}$ represents the highest attack impacts on the robust aggregation rules in $\mathcal{S}$ while  $\sup_{j\in [q]}{|\langle Q_j, \bar{V}\rangle|}$ characterizes the highest attack impacts on the non-robust aggregation rules in $\mathcal{S}^{\prime}$.

\begin{theorem}[Convergence Analysis]\label{theorem:convergence}
Let assumptions~\ref{assum:unbiased},\ref{assum:bound_var},\ref{assum:bound_data_hetero} and \ref{assum:L-smoothness} hold, $K$ be the number of sampled clients in each round, $h_{m}$ be the maximal Byzantine client number in each of $T$ rounds of training, $\{\alpha_1, \ldots, \alpha_M\}$ be the robust coefficients for aggregation rules in $\mathcal{S}$ corresponding to $h_m$, $P$ be the sampling distribution over $\mathcal{S}$. If $h_{m} < K/2$, the learning rate $\eta$ satisfies that
\begin{footnotesize}
\begin{equation}\label{equ:learning_rate_momentum}
\begin{aligned}
   \eta = \min\left\{\sqrt{\frac{32 L (\mathcal{F}(x^{0}) - \mathcal{F}^{*})+(6+\frac{10}{K-h_m}) G_l^2 +7 G_g^2 }{(8LT)(80L(\frac{G_l^2}{K-h_m}+G_g^2)+ 240 L \mathbb{E}_{\mathcal{AGR}_{i}\sim p}[\alpha_i]G_l^2)}}, \frac{1}{8L}\right\},
\end{aligned}
\end{equation}
\end{footnotesize}
and the momentum parameter $\beta = 1-8L\eta$, then it holds that
\begin{footnotesize}
\begin{equation}\label{equ:convergence}
\begin{aligned}
    &\frac{1}{T}\sum_{t=1}^{T}\mathbb{E}\|\nabla\mathcal{F}(x^{t-1})\|^2
    \leq \frac{32L(\mathcal{F}(x^{0}) - \mathcal{F}^{*})}{T} +  \frac{\frac{6}{K-h_m} G_l^2 + 3G_g^2}{T}+ \frac{2\|\nabla \mathcal{F}(x^{0})\|^2}{T}+ 15\mathbb{E}_{\mathcal{AGR}_{i}\sim p}[\alpha_i]G_g^2\\
    &+\sqrt{32 L (\mathcal{F}(x^{0}) - \mathcal{F}^{*})+(6+\frac{10}{K-h_m}) G_l^2 +7 G_g^2}\cdot\sqrt{\frac{640 L^2(\frac{G_l^2}{K-h_m}+G_g^2)+ 1920 L^2 \mathbb{E}_{\mathcal{AGR}_{i}\sim p}[\alpha_i]G_l^2}{T}}. 
\end{aligned}
\end{equation}
\end{footnotesize}
\end{theorem}

Theorem~\ref{theorem:convergence} states that the dynamic defense strategy with an appropriate time-varying learning rate converges to a neighborhood of a first order stationary point in expectation. The radius of this neighborhood (i.e., $15\mathbb{E}_{\mathcal{AGR}_{i}\sim p}[\alpha_i]G_g^2$) depends on both the data heterogeneity and the expected robust level. Note that the $h_m$ malicious clients in each round affect the error bound through the term of $\mathbb{E}_{\mathcal{AGR}_{i}\sim p}[\alpha_i]$. As stated in the theorem, the robust level of each aggregation rule in the candidate set is coupled with $h_m$. When $h_m \rightarrow 0$, each $\alpha_i$ also tends to be a small value that is only determined by the aggregation bias induced by $\mathcal{AGR}_i$.
Given the data heterogeneity, the error bound can be reduced by decreasing $\mathbb{E}_{\mathcal{AGR}_{i}\sim p}[\alpha_i]$, which can be achieved by assigning higher sampling weights to the robust aggregation rules. When $G_g^2= 0$, the convergence rate can be simplified to $\mathcal{O}(\frac{G_l^2}{\sqrt{T}}\cdot\sqrt{\mathbb{E}_{\mathcal{AGR}_{i}\sim p}[\alpha_i]+\frac{c}{K-h_m}})$, matching convergence rate in data homogeneous settings~\cite{karimireddy2021learning}.

Furthermore, we present two representative sampling strategies for instantiating the dynamic defense strategy, including {\it uniform sampling} and {\it weighted sampling}, which serve as case studies for a better understanding of the Byzantine robustness provided by a black-box server.
Specifically, when the server lacks prior knowledge of attacks, a commonly employed practice is uniform sampling. During each round of training, the server randomly samples an aggregation rule from the candidate set $\mathcal{S}$ following the uniform probability distribution $P = [\frac{1}{M}, \ldots, \frac{1}{M}]$.

In cases where the server has prior knowledge, e.g., an approximate model update derived from a root dataset $\xi^0$~\cite{cao2020fltrust,xie2020zeno++}, it becomes feasible to optimize the sampling strategy. For instance, during the $t$-th round of training, the server fine-tunes the global model $x_{t}$ on $\xi^0$ to produce a trustable model update $\Delta_0$. After that, upon receiving model updates from benign and malicious clients, the server employs every aggregation rule in $\mathcal{S}$ to combine these updates, and adopts a weighted sampling strategy: $\mathcal{R} = \{\mathcal{R}_{j}, \mathcal{AGR}_j\in \mathcal{S} \}$, where $\mathcal{R}$ denotes the aggregated results of all $M$ rules, with each update $\mathcal{R}_j$ being sampled based on a probability $p_j$ that is proportional to its similarity with the trustable model update.
Section~\ref{sec:exp} contains some empirical studies of these two sampling strategies.

\subsection{White-box Servers with Dynamic Defense v.s. Black-box Servers with Dynamic Defense}\label{subsec:thought_experiment}

As discussed in Section~\ref{subsec:dds}, an inaccessible candidate set $\mathcal{S}$ of aggregation rules is a crucial characteristic of a black-box server in the dynamic defense strategy. 

However, a white-box server can also achieve similar unpredictability with an accessible candidate set of aggregation rules, as pointed out by previous study~\cite{ramezani2022mixtailor}. Here we provide discussions on these two settings to highlight the importance of a black-box server and analyze the extent to which the black-box server can enhance the robustness.

For the white-box server setting where the candidate set $\mathcal{S}$ is accessible to clients, malicious clients can easily conduct local experiments to quantify attack impact $\alpha_{i,j}$ of each attack $\mathcal{A}_i \in \mathcal{A}$ to each aggregation rule $\mathcal{AGR}_j \in \mathcal{S}$.
As a result, malicious clients are able to choose the attack method that can incur the highest attack impacts considering all the aggregation rules to conduct continuous attacks. 
Formally, let $P_d$ be the defense strategy, i.e., the sampling distribution over $\mathcal{S}$, adopted by the server, the attack impact in each round can be characterized by $\max_{i\in \mathcal{A}}\mathbb{E}_{j\sim P_d}[\alpha_{i,j}]$.

However, when the server maintains an inaccessible set of aggregation rules, i.e., the black-box server setting,  
no matter what attack strategy is adopted by malicious clients, the expected attack impact on the FL system would not be higher than $\max_{i\in \mathcal{A}}\mathbb{E}_{j\sim P_d}[\alpha_{i,j}]$. Let $P_a$ be the attack strategy adopted by malicious clients, it should hold that:
\begin{align*}
    \mathbb{E}_{i\sim P_a}\mathbb{E}_{j\sim P_d}[\alpha_{i,j}] \leq \max_{i\in \mathcal{A}}\mathbb{E}_{j\sim P_d}[\alpha_{i,j}].
\end{align*}

In summary, although a certain level of unpredictability can be provided by random sampling from the multiple aggregation rules, a white-box server always provides prior knowledge to malicious clients for finding an optimal attack strategy. 
Such insights further highlight that \textit{the black-box property can improve Byzantine robustness since it takes the server to an equal position with the hidden malicious clients in the game of defense and attack}.

\section{Experiments}
\label{sec:exp}
We conduct a series of experiments to provide empirical observations on the robustness of FL systems with a black-box server that defends against various Byzantine attacks.

\subsection{Experimental Settings}\label{subsec:settings}
\paragraph{Datasets and Models}  
We federally train \texttt{convnet2} model on FEMNIST~\cite{caldas2018leaf} and \texttt{VGG11}~\cite{simonyan2014very} model on CIFAR-10~\cite{krizhevsky2009learning}, where the CIFAR-10 dataset is split according to a Dirichlet distribution with parameter $0.5$. 

We build up all the experiments based on FederatedScope~\cite{xie2023federatedscope}, a developer-friendly FL platform.
The total client number is set to 200, and the client sampling ratio in each round is set to 20\%. We employ SGD as the optimizer, with the momentum parameters for FEMNIST and CIFAR-10 are 0 and 0.9 respectively. The proportion of malicious clients varies from 2.5\% to 10.0\%.

\begin{figure*}[t]
	\centering
    	\includegraphics[width=1.0\textwidth]{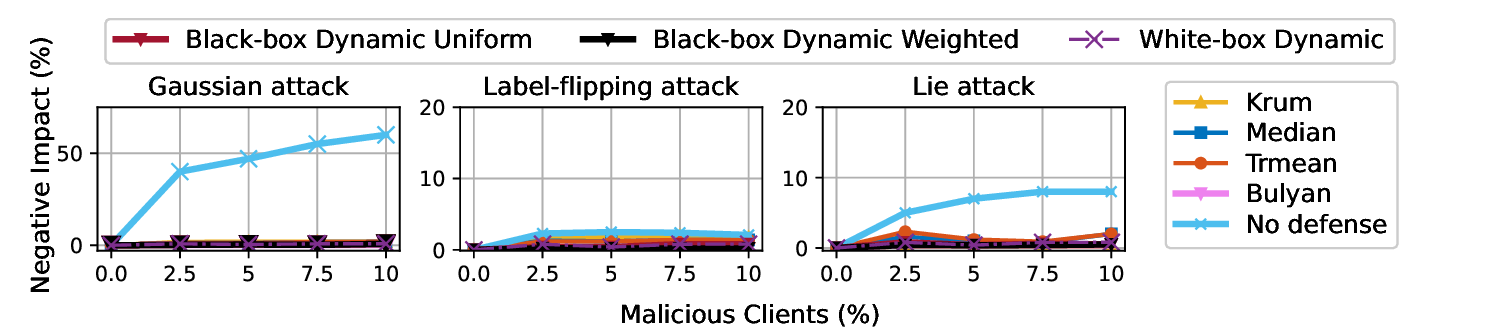}
    \caption{Defending against \texttt{AGR-agnostic} attacks on FEMNIST. \label{fig:general_attack_femnist}}
\end{figure*}

\paragraph{Aggregation Rules and Attack Algorithms}

For comparisons, we adopt four widely-used aggregation rules as the candidate defense algorithms, including Krum~\cite{blanchard2017machine}, Median~\cite{yin2018byzantine}, Trimmedmean~\cite{yin2018byzantine}, and Bulyan~\cite{guerraoui2018hidden}. Meanwhile, we resort to three \texttt{AGR-agnostic} attacks, i.e., Gaussian attack, label flipping and Lie attack~\cite{baruch2019little}, and two effective \texttt{AGR-adaptive} attacks, i.e., Fang attack~\cite{fang2020local} and She attack~\cite{shejwalkar2021manipulating} to conduct experiments on the Byzantine robustness. More details of the adopted aggregation rules and attack algorithms can be found in Appendix~\ref{subsec:aggre_rules}.

\paragraph{Compared Defending Strategies} 
For the white-box server, there are two strategies: (i) Applying a deterministic aggregation rule that is accessible to malicious clients (i.e., Krum, Median, Trimmedmean and Bulyan), and (ii) Dynamically sampling an aggregation rule from an accessible candidate set in each training round (as discussed in Section~\ref{subsec:thought_experiment}), denoted as ``White-box Dynamic'' in the figures. For the black-box server, we consider the two dynamic defense strategies that are discussed in Section~\ref{subsec:dds}, denoted as ``Black-box Dynamic Uniform'' and ``Black-box Dynamic Weighted''.

\paragraph{Evaluation Metric} We adopt the {\it negative impact}~\cite{shejwalkar2022back,fang2020local,shejwalkar2021manipulating} brought by Byzantine attacks as the evaluation metric, denoted as $I_{\theta}$. Denote $A_{ini}$ as the performance (e.g., accuracy) of the global model learned with \texttt{Fedavg} without being attacked, $A_{att}$ as the performance when there exist Byzantine attacks. The negative impact is calculated by: $I_{\theta}=\max (0,A_{ini}-A_{att})$.

\subsection{Results and Analysis} 

\paragraph{\texttt{AGR-agnostic} Attacks}
When the adopted aggregation rules are inaccessible, it is a reasonable and effective solution for malicious clients to perform \texttt{AGR-agnostic} attacks. Therefore we conduct experiments to study the Byzantine robustness of FL systems with a black-box server against \texttt{AGR-agnostic} attacks. The experimental results are demonstrated in Figure~\ref{fig:general_attack_femnist}. We report the negative impact brought by three \texttt{AGR-agnostic} attacks when the server applies static (i.e. Krum, Median, Trimmedmean and Bulyan) and dynamic defense strategies (i.e., ``White-box Dynamic'', ``Black-box Dynamic Uniform'' and ``Black-box Dynamic Weighted'') on FEMNIST dataset. From the figures, we can observe that both static and dynamic defense strategies can achieve competitive performance in defending all the adopted attack algorithms, including Gaussian attack, Label flipping, and Lie attack. The similar results on CIFAR-10 dataset are shown in Figure~\ref{fig:general_attack_cifar10} in Appendix~\ref{sec:add_exp}.

Further, as the proportion of malicious clients increases, the negative impacts brought by Byzantine attacks on the models learned without defense become larger, while the performance of models learned with both static and dynamic defense strategies stays at the same level. 
It is worth pointing out that {\tt Fedavg} (i.e., ``No defense'' in the figures) achieves a certain level of Byzantine robustness against the label flipping attack on FEMNIST dataset, which is consistent with previous study~\cite{shejwalkar2022back}. Such robustness is attributed to the client sampling procedure, since some \texttt{AGR-agnostic} attacks rely on a large amount of malicious clients and the attack continuity, while the server only samples a subset of clients (including both malicious and benign clients) in each FL training round.

\paragraph{\texttt{AGR-adaptive} Attacks}
\begin{figure*}[t]
	\centering
	\includegraphics[width=\textwidth]{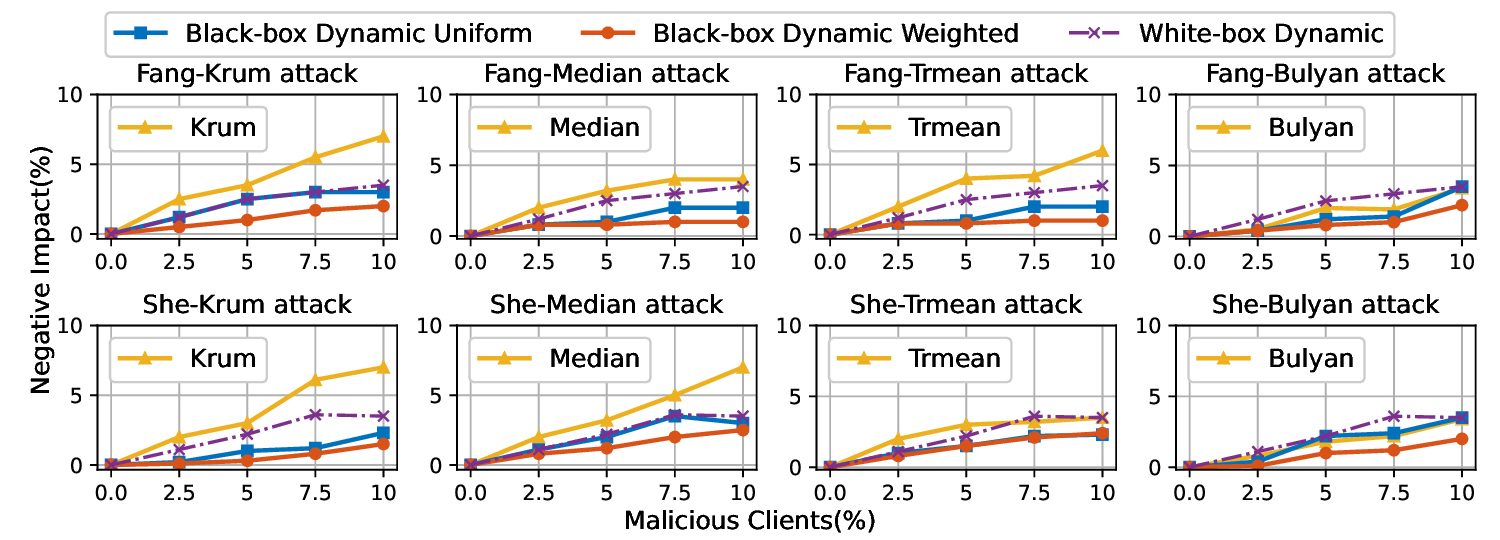}
    \caption{Defending against \texttt{AGR-adaptive} attacks on FEMNIST.\label{fig:adaptive_attack_femnist}}
\end{figure*}
\begin{figure*}[t]
	\centering
	\includegraphics[width=\textwidth]{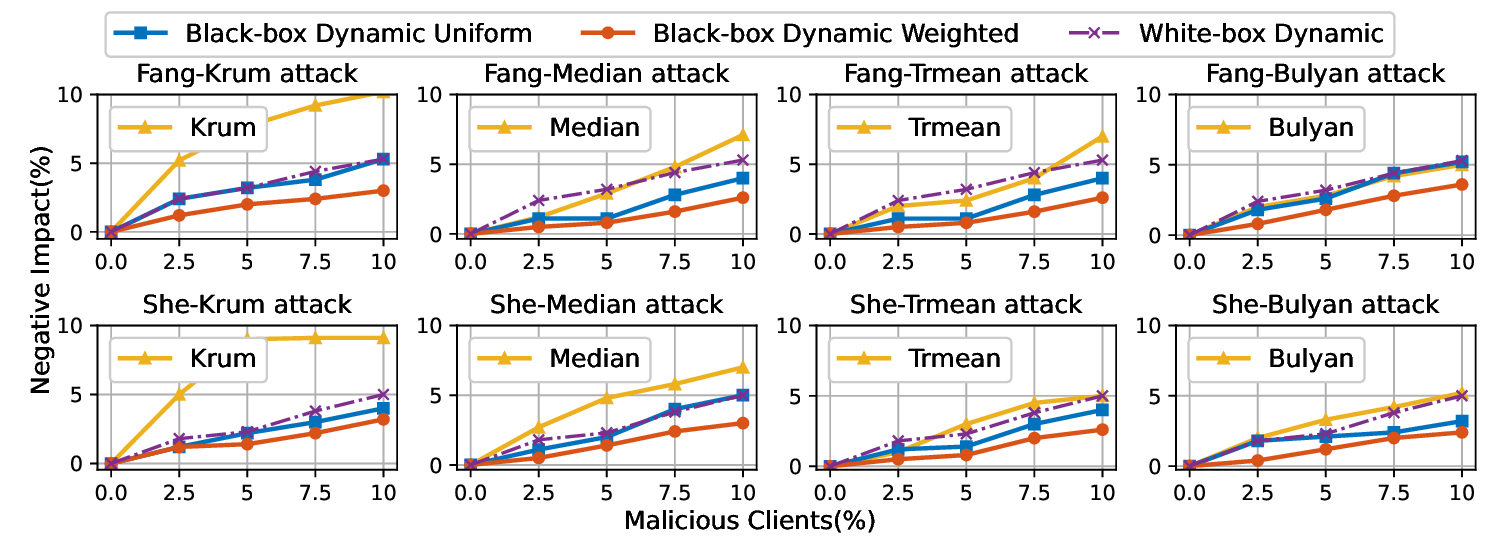}
    \vspace{-0.08in}
    \caption{Defending against \texttt{AGR-adaptive} attacks on CIFAR-10.\label{fig:adaptive_attack_cifar10}}
\end{figure*}

\begin{figure*}[t]
	\centering
	\includegraphics[width=\textwidth]{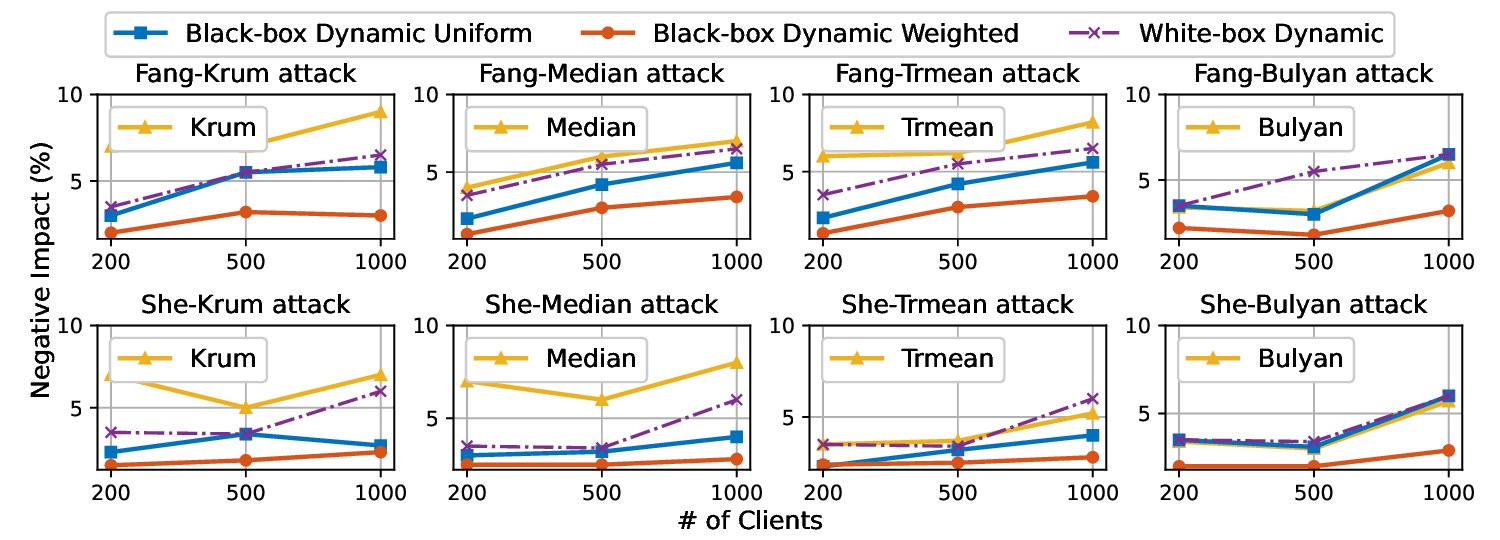}
    \caption{Negative impact caused by \texttt{AGR-adaptive} attacks w.r.t. the number of clients.\label{fig: exc_num_adaptive_attack_femnist}}
\end{figure*}

\begin{figure*}[t]
	\centering
	\includegraphics[width=\textwidth]{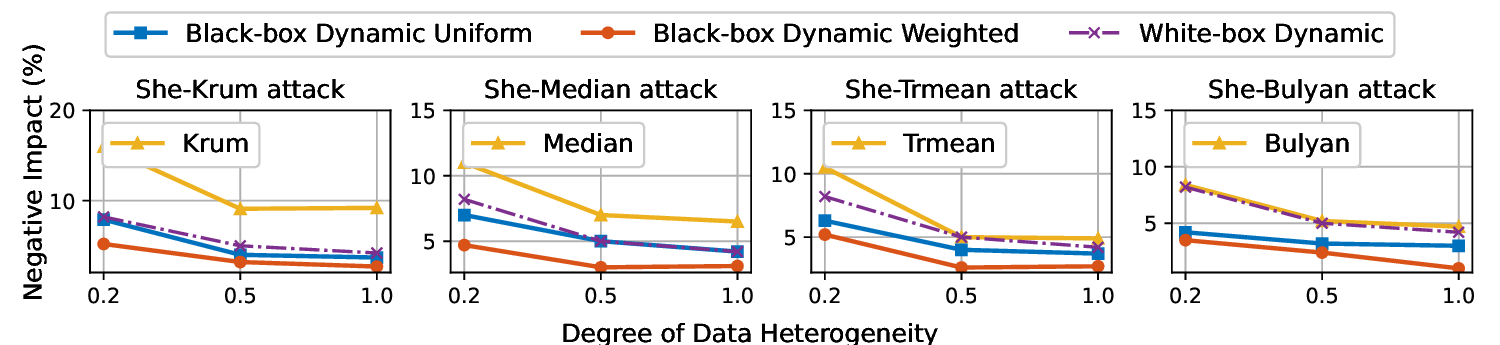}
    \caption{Negative impact caused by \texttt{AGR-adaptive} attacks w.r.t. the degree of data heterogeneity.\label{fig:noniid_adaptive_attack_cifar10}}
\end{figure*}

Malicious clients can apply \texttt{AGR-adaptive} attacks to incur the worst-case negative impact on model utility when the aggregation rules are accessible.
We conduct experiments on FEMNIST and CIFAR-10 datasets to compare the negative impact on FL systems with a white-box server and a black-box server applying dynamic defense strategies, as shown in Figure~\ref{fig:adaptive_attack_femnist} and \ref{fig:adaptive_attack_cifar10}. For the white-box server, we consider both static and Dynamic (``White-box Dynamic'') defense strategies. For both strategies, malicious clients tend to choose a specific attack algorithm according to the knowledge of the attack impacts. For example, when the server applies Krum, malicious clients adopt Fang attack or She attack targeting Krum, denoted as ``Fang-Krum attack'' or ``She-Krum attack'' in the figures, respectively. When the server samples aggregation rules from $\{$Krum, Median, Trimmedmean, 
Bulyan$\}$, malicious clients adopt the attack that can incur the highest attack impact from the attack set, e.g., $\{$Fang-Krum, Fang-Median, Fang-Trmean, 
Fang-Bulyan$\}$ for the Fang attack. We set up a black-box server that applies the two dynamic defense strategies discussed in Section~\ref{subsec:dds}, i.e., ``Black-box Dynamic Uniform'' and ``Black-box Dynamic Weighted'', for comparison.

From the figures, we observe that a black-box server with dynamic defense significantly outperforms the white-box server in defending against \texttt{AGR-adaptive} attacks. In particular, ``Black-box Dynamic Uniform'' shows superiority to both static and dynamic strategies adopted by the white-box server, and ``Black-box Dynamic Weighted'' presents further advantage to ``Black-box Dynamic Uniform'', which validates the advantage of black-box server in improving the Byzantine robustness. Although applying dynamic defense might also be successfully attacked by \texttt{AGR-adaptive} attacks when the server happens to choose the targeting aggregation rule in some FL training rounds, the negative impact is limited. The reason is that the dynamic of the aggregation rules would discontinue the contributions of malicious model updates and cause the effect of attacks intermittent. Such phenomena are further amplified by the client sampling procedure in FL.
Besides, the randomness and inaccessibility of the candidate set in the black-box server enhance the unpredictability of the defense strategy, which preserves the server's equal position with hidden malicious clients in the game of attack and defense.

The superiority of ``Black-box Dynamic Weighted'' to ``Black-box Dynamic Uniform'' demonstrates a huge potential that the black-box server can take to an FL system in improving the Byzantine robustness. 
The ``Black-box Dynamic Uniform'' essentially presents the worst-case defense effectiveness of a black-box server, and various prior knowledge can further take the black-box server to an advantageous position to the hidden malicious clients in the game of attack and defense.

\paragraph{Byzantine Robustness w.r.t. Clients Number}
To further understand the Byzantine robustness brought by a black-box server, we vary the total client number in $\{200, 500, 1000\}$ and fix the proportion of malicious clients as $10\%$. The experimental results are summarized in Figure~\ref{fig: exc_num_adaptive_attack_femnist}. 
From these results, we can conclude that a black-box server which adopts a dynamic defense strategy effectively defends against various Byzantine attacks and outperforms white-box servers which adopt deterministic or dynamic defense strategies. Thus, the advances of Byzantine robustness of a black-box server are stable w.r.t. the total number of clients.

\paragraph{Byzantine Robustness w.r.t. Data Heterogeneity}
We vary the degree of data heterogeneity in $\{0.2, 0.5, 1.0\}$ to synthesize the distributed CIFAR-10 dataset, where the degree is the parameter of the Dirichlet distribution used to split the dataset.
The experimental results are summarized in Figure~\ref{fig:noniid_adaptive_attack_cifar10}. 
From these results, we can also observe the stable advances of black-box server settings. 
 
\section{Related Works}
\label{sec:related_works}

\paragraph{Robust Aggregation Rules}
Recent studies on robust aggregation rules can be roughly divided into three categories: distance-based rules~\cite{blanchard2017machine,fung2018mitigating,wan2021shielding,wan2022shielding,cao2019understanding}, statistic-based rules~\cite{farhadkhani2022byzantine,guerraoui2018hidden,karimireddy2021byzantine,liu2023byzantine,pillutla2022robust,xie2018generalized}, and performance-based rules~\cite{cao2020fltrust,xie2019zeno,li2019abnormal,cao2019distributed}. 
Distance-based rules discriminate the Byzantine attacks via comparing the distance between the received model updates and then filter out the outliers. For example,  Krum~\cite{blanchard2017machine} sorts the Euclidean distances between updates and selects only updates which are closest to their neighbors. FoolsGold~\cite{fung2018mitigating} computes the cosine similarity between the updates to determine their contributions in the aggregation. 
Statistic-based rules exploit some robust statistics to perform aggregation to improve the robustness. For example, Median and Trimmedmean~\cite{yin2018byzantine} take the coordinate-wise median and the coordinate-wise trimmed mean over model updates as the aggregated result. RESAM~\cite{farhadkhani2022byzantine} demonstrates that the momentum of gradients can help improve the robustness of aggregation. Performance-based rules validate the updates with the help of a clean dataset. For example, FLtrust~\cite{cao2020fltrust} assigns trust scores for model updates based on the distance from a reliable model update computed on a small clean dataset. Zeno~\cite{xie2019zeno} computes scores based on its magnitude and the descendant of the loss function on a small validation set. Besides, there is a line of research demonstrating that robustness can be improved by preprocessing the model updates. e.g., gradients bucketing~\cite{karimireddy2021byzantine} or splitting~\cite{liu2023byzantine}, before applying the aggregation rules in non-IID settings.

\paragraph{Randomization Enhanced Robustness}
There has been a line of literature~\cite {lecuyer2019certified,cohen2019certified,pinot2020randomization,meunier2021mixed, ramezani2022mixtailor} building the connection between randomization and adversarial robustness of deep learning. The study in \cite{lecuyer2019certified} first introduces the differential private~\cite{dwork2006differential,zhao2020latent,ren2018LoPub} randomness into the deep learning procedure, which guarantees any small changes incurred by adversarial examples wouldn't produce an overwhelmingly negative impact on the model performance. Further, researchers~\cite{pinot2020randomization} prove that any deterministic classifier can be outperformed by a randomized one when evaluated against deterministic attack strategies, and consider the randomization for both the classifier and the attacker from a game theoretic point of view and confirms that the use of randomization can enhance adversarial robustness~\cite{meunier2021mixed}. 
Inspired by previous studies that focus on the robustness of deep learning to adversarial examples, we discuss the robustness gain brought by a black-box server against Byzantine updates.
\section{Conclusions}
\label{sec:conclu}

In this paper, we conduct the first study on Byzantine robustness of the FL system with a black-box server. In particular, we set up an \texttt{attack and defense} model to compare the Byzantine robustness between the white-box and black-box server settings, applying both the static and dynamic defense strategies. We provide theoretical analysis for the dynamic defense strategies and accordingly analyze the extent to which a black-box server with a dynamic defense strategy can improve its robustness. The analysis shows that the black-box property can reduce the attack impact from a maximum level to an expectation level. We provide empirical evidence of the effectiveness of a black-box server in defending against advanced Byzantine attacks.

\small
\bibliographystyle{ACM-Reference-Format}
\bibliography{reference}
\newpage
\appendix
\onecolumn

\appendix
\onecolumn

\section{\texttt{Fedavg}}
\label{sec:fedavg}

Here we provide the pseudocode of Federated Averaging algorithm (\texttt{Fedavg}). As shown in Algorithm~\ref{alg:fedavg}, in $t$-th round of training, the server first samples a set $C^t$ of active clients to participate the training, and broadcasts the current model $x^t$ to those clients. Upon receiving $x^t$, each client in $C^t$ trains $x^t$ on his/her local data and computes the local update. Together with the local data size, the local update is sent to the central server. After receiving all the local updates, the server takes weighted average over the local updates based on received local data sizes. The result is taken as the final update and used to update the current model $x^t$. The process proceeds for $T$ rounds to ensure the model convergence. 

\begin{algorithm}[H]
    \caption{{\tt Fedavg}}
    \label{alg:fedavg}
    \KwIn {federated dataset $\{\xi_1, \xi_2\, ..., \xi_n\}$, learning rate $\eta$, round number $T$}
    \KwOut{the final global model $x^{T}$}
    {$x_{0} \leftarrow$ Initialize the model parameter at the server.}\\
    \For{$t = 0$ to $T$}{ 
        {Server samples a set of $K$ active clients $C^t$\, and sends $x_{t}$ to each client in $C^t$.}\\
        \For{each client $c_i\in C^t$}{
            {Sets $x^{t}$ as the model parameter.}\\
            {Trains $x^{t}$ on the sampled data $\xi^t_i\sim \xi_i$}.\\
            {Sends the model update $\Delta x^t_i$ and local data size $|\xi_i|$ to the server.}
        }
    $x^{t+1}\leftarrow x^{t}+\eta\sum_{c_i\in C^t}\frac{|\xi_i|}{\sum_{i}|\xi_i|}\Delta x^t_i$.\\
    } 
\end{algorithm}

\section{Detailed Derivations}
\label{sec:proof}

In this section, we present the detailed derivations of the robustness (Theorem~\ref{theorem:robust} in the main paper) and convergence analysis (Theorem~\ref{theorem:convergence}) of the proposed \textit{dynamic defense strategy} (DDS) respectively. 

\subsection{Proof of Theorem~\ref{theorem:robust}}

\begin{proof}
We validate the two conditions in the definition of Byzantine robustness (shown in Section~\ref{sec:preliminary}) respectively. 

Regarding condition (i), according to the expectation formula, when given a probability distribution satisfying the condition shown in Equation~\eqref{equ:pro_mass}, it holds that
\begin{equation}
\begin{aligned}
    &\mathbb{E}_{\mathcal{AGR}_i\in \mathcal{S}}[\langle Q_i, \bar{V} \rangle]  = \sum_{j\in [q]}p_j\cdot\langle Q_j, \bar{V}\rangle + \sum_{i\notin [q]}p_i\cdot\langle Q_i, \bar{V}\rangle \\
    & > \frac{\inf_{i\notin [q]}{\langle Q_i, \bar{V}\rangle}}{\sup_{j\in [q]}{|\langle Q_j, \bar{V}\rangle|} + \inf_{i\notin [q]}{\langle Q_i, \bar{V}\rangle}}\cdot \inf_{j\in [q]}{\langle Q_j, \bar{V}\rangle} + \frac{\sup_{j\in [q]}{|\langle Q_i, \bar{V}\rangle|}}{\sup_{j\in [q]}{|\langle Q_j, \bar{V}\rangle|} + \inf_{i\notin [q]}{\langle Q_i, \bar{V}\rangle}}\cdot \inf_{i\notin [q]}{\langle Q_i, \bar{V}\rangle} \\
    & = - \frac{\inf_{i\notin [q]}{\langle Q_i, \bar{V}\rangle}}{\sup_{j\in [q]}{|\langle Q_j, \bar{V}\rangle|} + \inf_{i\notin [q]}{\langle Q_i, \bar{V}\rangle}}\cdot \sup_{j\in [q]}{|\langle Q_i, \bar{V}\rangle|} + \frac{\sup_{j\in [q]}{|\langle Q_i, \bar{V}\rangle|}}{\sup_{j\in [q]}{|\langle Q_j, \bar{V}\rangle|} + \inf_{i\notin [q]}{\langle Q_i, \bar{V}\rangle}}\cdot \inf_{i\notin [q]}{\langle Q_i, \bar{V}\rangle}  = 0
\end{aligned}
\end{equation}
Regarding condition (ii), suppose that the robust level of $\mathcal{AGR}_i$ is $(h, \alpha_i)$, then it holds that
\begin{equation}
\begin{aligned}
    &\mathbb{E}_{\mathcal{AGR}_i\in \mathcal{S}}[\|Q_i - \bar{V}\|^2] \\
    & = \sum_{i}p_i\cdot\|Q_i - \bar{V}\|^2 \\
    & \leq \sum_{i}p_i \cdot \frac{\alpha_i}{|V|}\sum_{i\in V}\|V_i-\bar{V}\|^2\\
    & = \frac{\mathbb{E}_{\mathcal{AGR}_{i}\sim p}[\alpha_i]}{|V|}\sum_{i\in V}\|V_i-\bar{V}\|^2
\end{aligned}
\end{equation}

For now, the two conditions of Byzantine robustness have been validated and the theorem has been proved. 
\end{proof}

\subsection{Proof of Theorem~\ref{theorem:convergence}}
We basically follow the proof of \cite{karimireddy2021byzantine}, with main differences in the definition of Byzantine robustness, aggregation strategy, and assumption of the data heterogeneity.  
\begin{proof}

Without loss of generality, consider the $t$-th aggregation, the update of the model parameter $x^{t-1}$ can be formulated as follows:
$$
    x^{t} = x^{t-1} - \eta m^t, 
$$
where $m^t$ is the aggregated momentum. For each client $c_i$, the local momentum is computed by 
$$m^t_i =  (1-\beta) m_{i}^{t-1} + \beta\nabla f_i(x^{t-1};\xi^{t-1}),$$ where $m^1_i = \nabla f_i(x^{0};\xi^{0})$.  Let $\eta \leq \frac{1}{L}$, Combining the randomness in the training process and the \textit{L}-smoothness property of the loss function $\mathcal{F}$ shown in Assumption~\ref{assum:L-smoothness}, we can deduce a descent inequality: 
\begin{equation}\label{equ:descent_formula}
\begin{aligned}
    \mathbb{E}\mathcal{F}(x^{t})
    &\leq \mathcal{F}(x^{t-1}) - \eta\mathbb{E}\langle \nabla \mathcal{F}(x^{t-1}), m^{t}\rangle + \frac{L}{2}\eta^{2}\mathbb{E}\Vert m^{t}\Vert^{2}\\
    &\leq \mathcal{F}(x^{t-1}) - \eta\mathbb{E}\langle \nabla \mathcal{F}(x^{t-1}), m^{t}\rangle + \frac{\eta}{2}\mathbb{E}\Vert m^{t}\Vert^{2}\\
    &\leq \mathcal{F}(x^{t-1}) + \frac{\eta}{2}\mathbb{E}\Vert m^{t} - \nabla \mathcal{F}(x^{t-1})\Vert^{2} - \frac{\eta}{2}\mathbb{E}\Vert \nabla \mathcal{F}(x^{t-1})\Vert^{2}\\
    &= \mathcal{F}(x^{t-1}) + \frac{\eta}{2}\underbrace{\mathbb{E}\Vert m^{t} +\bar{m}^t-\bar{m}^t- \nabla \mathcal{F}(x^{t-1})\Vert^{2}}_{\mathbb{E}\|\sum_{i\leq n}{a_i}\|^2\leq n\sum_{i\leq n}\mathbb{E}\|a_i\|^2} - \frac{\eta}{2}\mathbb{E}\Vert \nabla \mathcal{F}(x^{t-1})\Vert^{2}\\
    &\leq \mathcal{F}(x^{t-1}) + \eta\mathbb{E}\|\underbrace{\bar{m}^t- \nabla \mathcal{F}(x^{t-1})}_{=:\bar{e}^t}\|^2+\eta\mathbb{E}\Vert m^{t} -\bar{m}^t\Vert^{2} - \frac{\eta}{2}\mathbb{E}\Vert \nabla \mathcal{F}(x^{t-1})\Vert^{2},
\end{aligned}
\end{equation}
where $\bar{m}^t$ denotes the average momentum, $\mathbb{E}[\cdot]$ accounts for all the randomness in the current round of training. $\mathbb{E}[\cdot]$ can be replaced by $\mathbb{E}_{\alpha, \xi^{t-1}, V'}[\cdot]$ because the randomness in the current round can be summarized in three folds: i) the sampled data $\xi^{t-1} = \{\xi^{t-1}_1,\ldots,\xi^{t-1}_n,\ldots,\xi^{0}_1,\ldots,\xi^{0}_n\}$; ii) updates $V' = \{B, V\}$ provided by sampled clients $C^t$; iii) the sampled aggregation rule $\mathcal{AGR}_{\alpha}$. 

We first bound the \textbf{aggregation error} $\mathbb{E}_{\xi^{t-1}}\Vert m^{t} -\bar{m}^t\Vert^{2}$. Here we reuse inner results in the proof of \cite{karimireddy2021byzantine}, that is

\begin{equation}\label{equ:aggregation_error}
\begin{aligned}
    \mathbb{E}_{\xi^{t-1}}\|m_i^t - \bar{m}^t\|^2 \leq (6\beta G_l^2+3G_g^2)+(6 G_l^2 - 3G_g^2)(1-\beta)^t.
    \end{aligned}
\end{equation}

Let $C^t_b \subset C^t$ denotes the participating benign clients in the $t$-th round and $h_t$ denote the number of participating malicious clients in the $t$-th round. Then according to the definition of the Byzantine robustness, we can derive that
\begin{equation*}
\begin{aligned}
    \mathbb{E}_{\alpha, \xi^{t-1}, V'}\|m^{t} - \bar{m}^t\|^2 &\leq \frac{\mathbb{E}_{\mathcal{AGR}_{\alpha}\sim p}[\alpha]}{K-h_{t}} \sum_{c_i\in C^t_b}\mathbb{E}_{\xi^{t-1}}\|m_i^{t} - \bar{m}^t\|^2 \\
    &\leq \mathbb{E}_{\mathcal{AGR}_{\alpha}\sim p}[\alpha][(6\beta G_l^2+3G_g^2)+(6 G_l^2 - 3G_g^2)(1-\beta)^t.
\end{aligned}
\end{equation*}

Then we bound the \textbf{momentum deviation} $\mathbb{E}\|\bar{e}^t\|^2$. Let $\bar{\nabla f_i}(x^{t-1};\xi^{t-1}_i)= \frac{1}{|C^t|}\sum_{c_i\in C^t}\nabla f_i(x^{t-1};\xi^{t-1}_i)$, it holds that 
\begin{equation}\label{equ:error}
\begin{aligned}
    \mathbb{E}\|\bar{e}^t\|^2 
    &= \mathbb{E}_{\alpha, \xi^{t-1}, V'}\|\bar{m}^t- \nabla \mathcal{F}(x^{t-1})\|^2\\
    &= \mathbb{E}_{\alpha, \xi^{t-1}, V'}\|\beta\bar{\nabla f_i}(x^{t-1};\xi^{t-1}_i) +(1-\beta)\bar{m}^{t-1} - \nabla \mathcal{F}(x^{t-1})\|^2\\
    &= \mathbb{E}_{\alpha, \xi^{t-1}, V'}\|\beta\bar{\nabla f_i}(x^{t-1};\xi^{t-1}_i) - \beta\nabla F(x^{t-1}) + \beta\nabla F(x^{t-1}) +(1-\beta)\bar{m}^{t-1} - \nabla \mathcal{F}(x^{t-1})\|^2\\
    &\leq  \mathbb{E}_{\alpha, \xi^{t-1}, V'}\|\beta\nabla F(x^{t-1}) +(1-\beta)\bar{m}^{t-1} - \nabla \mathcal{F}(x^{t-1})\|^2+\beta^2\mathbb{E}_{\xi^{t-1}}\|\bar{\nabla f_i}(x^{t-1};\xi^{t-1}_i) - \nabla F(x^{t-1})\|^2 \\
    &\leq  (1-\beta)^2 \underbrace{\mathbb{E}_{\alpha, \xi^{t-1}, V'}\|(\bar{m}^{t-1} -\nabla F(x^{t-2}))+(\nabla F(x^{t-2})- \nabla F(x^{t-1}))\|^2}_{(a+b)^2\leq (1+\gamma)a^2 + (1+\frac{1}{\gamma})b^2} \\
    &\qquad+ \beta^2\mathbb{E}_{\xi^{t-1}}\|\bar{\nabla f_i}(x^{t-1};\xi^{t-1}_i) - \nabla F(x^{t-1})\|^2\\
    &\leq  (1-\beta)(1+\frac{\beta}{2}) \mathbb{E}_{\alpha, \xi^{t-1}, V'}\|\bar{m}^{t-1} -\nabla F(x^{t-2})\|^2+(1-\beta)(1+\frac{2}{\beta})\mathbb{E}_{\xi^{t-1}}\|\nabla F(x^{t-2})- \nabla F(x^{t-2})\|^2 \\
    &\qquad+ \beta^2\mathbb{E}_{\xi^{t-1}}\|\bar{\nabla f_i}(x^{t-1};\xi^{t-1}_i) - \nabla F(x^{t-1})\|^2\\
    &\leq  (1-\frac{\beta}{2}) \mathbb{E}_{\alpha, \xi^{t-1}, V'}\|\bar{e}^{t-1}\|^2+\frac{2L^2}{\beta}\mathbb{E}_{\xi^{t-1}}\|x^{t-2}- x^{t-1}\|^2 + \beta^2\underbrace{\mathbb{E}\|\bar{\nabla f_i}(x^{t-1};\xi^{t-1}_i) - \nabla F(x^{t-1})\|^2}_{L-smoothness}\\
    &\leq  (1-\frac{\beta}{2}) \mathbb{E}_{\alpha, \xi^{t-1}, V'}\|\bar{e}^{t-1}\|^2+\frac{2L^2\eta^2}{\beta}\mathbb{E}_{\xi^{t-1}}\|x^{t-2}- x^{t-1}\|^2 + \beta^2\mathbb{E}_{\xi^{t-1}}\|\bar{\nabla f_i}(x^{t-1};\xi^{t-1}_i) - \nabla F(x^{t-1})\|^2\\
    & \leq (1-\frac{\beta}{2})\mathbb{E}_{\alpha, \xi^{t-1}, V'}\|\bar{e}^{t-1}\|^2
    +\frac{6L^2\eta^2}{\beta}\|\bar{e}^{t-1} \|^2 + \frac{6L^2\eta^2}{\beta}\mathbb{E}_{\alpha, \xi^{t-1}, V'}\|m^{t-1}\|^2 \\
    &\qquad\qquad+ \frac{6L^2\eta^2}{\beta}\mathbb{E}_{\xi^{t-1}}\|\nabla \mathcal{F}(x^{t-2})\|^2 + \beta^2\mathbb{E}_{\xi^{t-1}}\|\bar{\nabla f_i}(x^{t-1};\xi^{t-1}_i) - \nabla F(x^{t-1})\|^2
\end{aligned}
\end{equation}

According to Assumption~\ref{assum:unbiased} and Assumption~\ref{assum:bound_data_hetero}, we can derive that
\begin{equation}\label{equ:error}
\begin{aligned}
&\mathbb{E}_{\xi^{t-1}}\|\bar{\nabla f_i}(x^{t-1};\xi^{t-1}_i) - \nabla F(x^{t-1})\|^2 \\
&= \mathbb{E}_{\xi^{t-1}}\left\|\frac{1}{K-h_t}\sum_{c_i\in C^t_b}\nabla f_i(x^{t-1};\xi^{t-1}_i) - \frac{1}{K-h_t}\sum_{c_i\in C^t_b}\nabla f_i(x^{t-1})+ \frac{1}{K-h_t}\sum_{c_i\in C^t_b}\nabla f_i(x^{t-1}) - \nabla F(x^{t-1})\right\|^2\\
&= \mathbb{E}_{\xi^{t-1}}\left[\left\|\frac{1}{K-h_t}(\sum_{c_i\in C^t_b}\nabla f_i(x^{t-1};\xi^{t-1}_i) - \sum_{c_i\in C^t_b}\nabla f_i(x^{t-1}))\right\|^2+ \left\|\frac{1}{K-h_t}\sum_{c_i\in C^t_b}\nabla f_i(x^{t-1}) - \nabla F(x^{t-1})\right\|^2\right]\\
&\leq \frac{G_l^2}{K-h_m}+G_g^2
\end{aligned}
\end{equation}

Let $64L^2\eta^2\leq \beta^2$, then it holds that

\begin{equation}\label{equ:error_bound}
\begin{aligned}
    \mathbb{E}\|\bar{e}^t\|^2 \leq (1-\frac{2\beta}{5})\mathbb{E}\|\bar{e}^{t-1}\|^2
    + \frac{\beta}{10}\mathbb{E}\|m^{t-1}-\bar{m}^{t-1}\|^2 + \frac{\beta}{10}\mathbb{E}\|\nabla \mathcal{F}(x^{t-2})\|^2 + \beta^2(\frac{G_l^2}{K-h_m}+G_g^2).
\end{aligned}
\end{equation}

Adding $\frac{5\eta}{2\beta}\mathbb{E}\|\bar{e}^t\|^2$ on both sides of Eq.~\eqref{equ:descent_formula} and expanding the term $\frac{5\eta}{2\beta}\mathbb{E}\|\bar{e}^t\|^2$ in the right side gives

\begin{equation}
\begin{aligned}
    \mathbb{E}\mathcal{F}(x^{t}) + \frac{5\eta}{2\beta}\mathbb{E}\|\bar{e}^t\|^2
    &\leq \mathcal{F}(x^{t-1}) + \eta\mathbb{E}\|\bar{e}^t\|^2+\eta\mathbb{E}\Vert m^{t} -\bar{m}^t\Vert^{2} - \frac{\eta}{2}\mathbb{E}\Vert \nabla \mathcal{F}(x^{t-1})\Vert^{2} + \frac{5\eta}{2\beta}\mathbb{E}\|\bar{e}^t\|^2\\
    & \leq \mathcal{F}(x^{t-1}) + \eta\mathbb{E}\|\bar{e}^t\|^2+\eta\mathbb{E}\Vert m^{t} -\bar{m}^t\Vert^{2} - \frac{\eta}{2}\mathbb{E}\Vert \nabla \mathcal{F}(x^{t-1})\Vert^{2} -\eta\mathbb{E}\|\bar{e}^{t-1}\|^2\\
    &\qquad\qquad+ \frac{5\eta}{2\beta}\mathbb{E}\|\bar{e}^{t-1}\|^2 + \frac{\eta}{4}\mathbb{E}\|m^{t-1}-\bar{m}^{t-1}\|^2 \\
    &\qquad\qquad\qquad\qquad+ \frac{\eta}{4}\mathbb{E}\|\nabla \mathcal{F}(x^{t-2})\|^2 + \frac{5\eta\beta}{2}(\frac{G_l^2}{K-h_m}+G_g^2)
\end{aligned}
\end{equation}

Substituting the aggregation error bound shown in Eq.~\eqref{equ:aggregation_error} and rearranging yields
\begin{equation}
\begin{aligned}\label{equ:recursion}
    &\underbrace{\mathbb{E}\mathcal{F}(x^{t}) - \mathcal{F}^{*} + (\frac{5\eta}{2\beta}-\eta)\mathbb{E}\|\bar{e}^t\|^2 + \frac{\eta}{4}\mathbb{E}\|\nabla \mathcal{F}(x^{t-1})\|^2}_{=:\epsilon^{t}}\\
    & \leq \underbrace{\mathbb{E}\mathcal{F}(x^{t-1}) - \mathcal{F}^{*} + (\frac{5\eta}{2\beta}-\eta)\mathbb{E}\|\bar{e}^{t-1}\|^2 + \frac{\eta}{4}\mathbb{E}\|\nabla \mathcal{F}(x^{t-2})\|^2}_{=:\epsilon^{t-1}} \\
    &\qquad\qquad- \frac{\eta}{4}\mathbb{E}\|\nabla \mathcal{F}(x^{t-1})\|^2 + \frac{5\eta\beta}{2}(\frac{G_l^2}{K-h_m}+G_g^2) \\
    &\qquad\qquad\qquad\qquad+\frac{5\eta}{4}\mathbb{E}_{\mathcal{AGR}_{\alpha}\sim p}[\alpha][(6\beta G_l^2+3G_g^2)+(6 G_l^2 - 3G_g^2)(1-\beta)^t] 
\end{aligned}
\end{equation}

According to the recursion of $\epsilon^t$, we can derive that
\begin{equation}
\begin{aligned}
    \epsilon^{1} &\leq \mathbb{E}\mathcal{F}(x^{1}) - \mathcal{F}^{*} + (\frac{5\eta}{2\beta}-\eta)\mathbb{E}\|\bar{e}^1\|^2 + \frac{\eta}{4}\mathbb{E}\|\nabla \mathcal{F}(x^{0})\|^2\\
    &\leq \mathcal{F}(x^{0}) - \mathcal{F}^{*} + (\frac{5\eta}{2\beta}-\eta)(\frac{G_l^2}{K-h_m}+G_g^2) + \frac{\eta}{4}\mathbb{E}\|\nabla \mathcal{F}(x^{0})\|^2\\
\end{aligned}
\end{equation}

Summing over $t$ from 2 to T on both sides of Eq.~\eqref{equ:recursion} yields
\begin{equation}
\begin{aligned}\label{equ:recursion_1}
    \sum_{t=2}^T \epsilon^{t}
    & \leq \sum_{t=2}^{T} \epsilon^{t-1} - \frac{\eta}{4}\sum_{t=2}^{T}\mathbb{E}\|\nabla \mathcal{F}(x^{t-1})\|^2 + \sum_{t=2}^{T}(\frac{5\eta\beta}{2}(\frac{G_l^2}{K-h_m}+G_g^2)) \\
    &\qquad\qquad+\sum_{t=2}^{T}\frac{5\eta}{4}\mathbb{E}_{\mathcal{AGR}_{\alpha}\sim p}[\alpha][(6\beta G_l^2+3G_g^2)+(6 G_l^2 - 3G_g^2)(1-\beta)^t] 
\end{aligned}
\end{equation}

Adding $\epsilon_1$ on both sides of Eq.~\eqref{equ:recursion_1} gives

\begin{equation}
\begin{aligned}\label{equ:recursion_2}
    \sum_{t=1}^T \epsilon^{t}
    & \leq \sum_{t=1}^{T-1} \epsilon^{t} - \sum_{t=1}^{T}\frac{\eta}{4}\mathbb{E}\|\nabla \mathcal{F}(x^{t-1})\|^2 +\frac{\eta}{2}\mathbb{E}\|\nabla \mathcal{F}(x^{0})\|^2+ \sum_{t=2}^{T}\frac{5\eta\beta}{2}(\frac{G_l^2}{K-h_m}+G_g^2)+ (\frac{5\eta}{2\beta}-\eta)(\frac{G_l^2}{K-h_m}+G_g^2)\\
    &\qquad+\sum_{t=2}^{T}\frac{5\eta}{4}\mathbb{E}_{\mathcal{AGR}_{\alpha}\sim p}[\alpha][(6\beta G_l^2+3G_g^2)+(6 G_l^2 - 3G_g^2)(1-\beta)^t] \\
    & \leq \sum_{t=1}^{T} \epsilon^{t} - \sum_{t=1}^{T}\frac{\eta}{4}\mathbb{E}\|\nabla \mathcal{F}(x^{t-1})\|^2 +\frac{\eta}{2}\|\nabla \mathcal{F}(x^{0})\|^2+ \sum_{t=2}^{T}\frac{5\eta\beta}{2}(\frac{G_l^2}{K-h_m}+G_g^2)+ (\frac{5\eta}{2\beta}-\eta)(\frac{G_l^2}{K-h_m}+G_g^2)\\
    &\qquad+\sum_{t=2}^{T}\frac{5\eta}{4}\mathbb{E}_{\mathcal{AGR}_{\alpha}\sim p}[\alpha][(6\beta G_l^2+3G_g^2)+(6 G_l^2 - 3G_g^2)(1-\beta)^t],
\end{aligned}
\end{equation}
where the second inequality holds since $\epsilon^{t}\geq 0$. $\mathbb{E}\|\nabla \mathcal{F}(x^{0})\|^2 = \|\nabla \mathcal{F}(x^{0})\|^2$ because the gradient of the initial point $x^{0}$ is deterministic. Then dividing both sides by $T$ and rearranging gives
\begin{equation}
\begin{aligned}
    &\frac{1}{T}\sum_{t=1}^{T}\mathbb{E}\|\nabla\mathcal{F}(x^{t-1})\|^2\\
    &\leq \frac{4(\mathcal{F}(x^{0}) - \mathcal{F}^{*})}{\eta T} + \frac{4}{T}\sum_{t=2}^{T}\left[\frac{5\beta}{2}(\frac{G_l^2}{K-h_m}+G_g^2)
    +\frac{5}{4}\mathbb{E}_{\mathcal{AGR}_{\alpha}\sim p}[\alpha][(6\beta G_l^2+3G_g^2)+(6 G_l^2 - 3G_g^2)\underbrace{(1-\beta)^t}_{\sum_{t}(1-\beta)^t\leq \frac{1}{\beta}}]\right] \\
    &\qquad+ \frac{1}{T}(\frac{10}{\beta}-4)(\frac{G_l^2}{K-h_m}+G_g^2) + \frac{2}{T}\mathbb{E}\|\nabla \mathcal{F}(x^{0})\|^2\\
    &\leq \frac{4(\mathcal{F}(x^{0}) - \mathcal{F}^{*})}{\eta T} + \frac{1}{T}\sum_{t=2}^{T}\left[10\beta(\frac{G_l^2}{K-h_m}+G_g^2)+ 30\beta \mathbb{E}_{\mathcal{AGR}_{\alpha}\sim p}[\alpha]G_l^2 \right] + \frac{1}{T}\sum_{t=2}^{T}\left[15 \mathbb{E}_{\mathcal{AGR}_{\alpha}\sim p}[\alpha]G_g^2\right] \\
    &\qquad\qquad+ \frac{6 G_l^2 - 3G_g^2}{\beta T}+ \frac{1}{T}(\frac{10}{\beta}-4)(\frac{G_l^2}{K-h_m}+G_g^2) + \frac{2}{T}\mathbb{E}\|\nabla \mathcal{F}(x^{0})\|^2\\
    &\leq \frac{4(\mathcal{F}(x^{0}) - \mathcal{F}^{*})}{\eta T} +  \frac{6 G_l^2 - 3G_g^2}{\beta T}+10\beta(\frac{G_l^2}{K-h_m}+G_g^2)+ 30\beta \mathbb{E}_{\mathcal{AGR}_{\alpha}\sim p}[\alpha]G_l^2 + 15\mathbb{E}_{\mathcal{AGR}_{\alpha}\sim p}[\alpha]G_g^2\\
    &\qquad\qquad+  \frac{1}{T}(\frac{10}{\beta}-4)(\frac{G_l^2}{K-h_m}+G_g^2) + \frac{2}{T}\mathbb{E}\|\nabla \mathcal{F}(x^{0})\|^2\\
    &\leq \frac{32 L(\mathcal{F}(x^{0}) - \mathcal{F}^{*})+(6+\frac{10}{K-h_m}) G_l^2 +7 G_g^2}{8L\eta T}+ \frac{2\mathbb{E}\|\nabla \mathcal{F}(x^{0})\|^2 - 4(\frac{G_l^2}{K-h_m}+G_g^2)}{T}\\
    &\qquad\qquad+80L\eta(\frac{G_l^2}{K-h_m}+G_g^2)+ 240 L\eta \mathbb{E}_{\mathcal{AGR}_{\alpha}\sim p}[\alpha]G_l^2  + 15\mathbb{E}_{\mathcal{AGR}_{\alpha}\sim p}[\alpha]G_g^2
\end{aligned}
\end{equation}
By taking 
$$
\eta = \min\left\{\sqrt{\frac{32 L (\mathcal{F}(x^{0}) - \mathcal{F}^{*})+(6+\frac{10}{K-h_m}) G_l^2 +7 G_g^2 }{(8LT)(80L(\frac{G_l^2}{K-h_m}+G_g^2)+ 240 L \mathbb{E}_{\mathcal{AGR}_{\alpha}\sim p}[\alpha]G_l^2)}}, \frac{1}{8L}\right\},
$$ 
we can derive that
\begin{equation}\label{equ:convergence}
\begin{aligned}
    \frac{1}{T}\sum_{t=1}^{T}\mathbb{E}&\|\nabla\mathcal{F}(x^{t-1})\|^2
    \leq \frac{32L(\mathcal{F}(x^{0}) - \mathcal{F}^{*})}{T} +  \frac{\frac{6}{K-h_m} G_l^2 + 3G_g^2}{T}+ \frac{2\|\nabla \mathcal{F}(x^{0})\|^2}{T}+ 15\mathbb{E}_{\mathcal{AGR}_{\alpha}\sim p}[\alpha]G_g^2\\
    &+\sqrt{32 L (\mathcal{F}(x^{0}) - \mathcal{F}^{*})+(6+\frac{10}{K-h_m}) G_l^2 +7 G_g^2}\cdot\sqrt{\frac{640 L^2(\frac{G_l^2}{K-h_m}+G_g^2)+ 1920 L^2 \mathbb{E}_{\mathcal{AGR}_{\alpha}\sim p}[\alpha]G_l^2}{T}} 
\end{aligned}
\end{equation}
For now, the theorem has been proved.
\end{proof}

\section{Additional Experimental Settings and Results}
\label{sec:add_exp}

\subsection{Aggregation rules and Attack algorithms}\label{subsec:aggre_rules}
Several widely-used robust aggregation rules are adopted as the candidate defense algorithms:
\begin{itemize}
\item{\bf Krum~\cite{blanchard2017machine}.} The server first computes the cumulative Euclidean distance between each update and its $m-h-2$ nearest neighbors, and then takes the mean of $k$ updates with the smallest cumulative Euclidean distances. In our experiments, $k=10$.
\item{\bf Median~\cite{yin2018byzantine}.} For each $j\in[d]$, the server sorts the $j$-th parameter of local updates and takes the median as the $j$-th parameter of the aggregated update. 
\item{\bf Trimmedmean~\cite{yin2018byzantine}.} For each $j$-th parameter, the server removes the largest and smallest $\beta$ of local updates, and takes the mean of the remaining. In our experiments, $\beta = 0.2$. Trimmedmean is also denoted by Trmean subsequently. 
\item{\bf Bulyan~\cite{guerraoui2018hidden}.} The server first selects $m-2h$ updates using Krum and then finds $m-4h$ parameters closest to the median for each coordinate, and finally computes the mean.
\end{itemize}

Meanwhile, the following attack algorithms are considered:
\begin{itemize}
\item{\bf Gaussian attack.} Malicious clients generate model updates by sampling from a Gaussian distribution $\mathcal{N}(0, \sigma^2)$. In our experiments, $\sigma = 0.5$. 
\item{\bf Label flipping.} Malicious clients train the model based on poisoned dataset in which each class label $c$ is flipped into $c_n-c$, where $c_n$ is the class number.
\item{\bf Lie attack~\cite{baruch2019little}.} Malicious clients compute the average $\overline{\Delta x}$ and standard deviation $\sigma$ of available benign updates and generate poisoned updates by $\Delta x = \overline{\Delta x} + z\cdot \sigma$, where $z$ is a scaling factor determined by the number of malicious and benign clients. 
\item{\bf Fang attack~\cite{fang2020local}.} 
For a specific aggregation rule $\mathcal{AGR}$, malicious clients compute the average $\overline{\Delta x}$ of the available benign updates and obtain a perturbation vector, $w  = -sign(\overline{\Delta x})$. Finally, malicious clients optimize to find a scaling factor $z$ so that the poisoned updates $\overline{\Delta x}+ z\cdot w$ can circumvent $\mathcal{AGR}$. 
\item{\bf She attack~\cite{shejwalkar2021manipulating}.} 
For a specific aggregation rule, malicious clients first compute the average $\overline{\Delta x}$ of the available benign updates and then generate poisoned updates by $\overline{\Delta x}+ z\cdot w$, where $z$ is a scaling factor that maximizes the attack impact, $w$ can be $\{ -sign(\overline{\Delta x}), -std(\overline{\Delta x}),  -\frac{\overline{\Delta x}}{\Vert\overline{\Delta x}\Vert}\}$.
\end{itemize}

\subsection{Additional Experimental Results}

\begin{figure}
	\centering
	\includegraphics[width=0.98\textwidth]{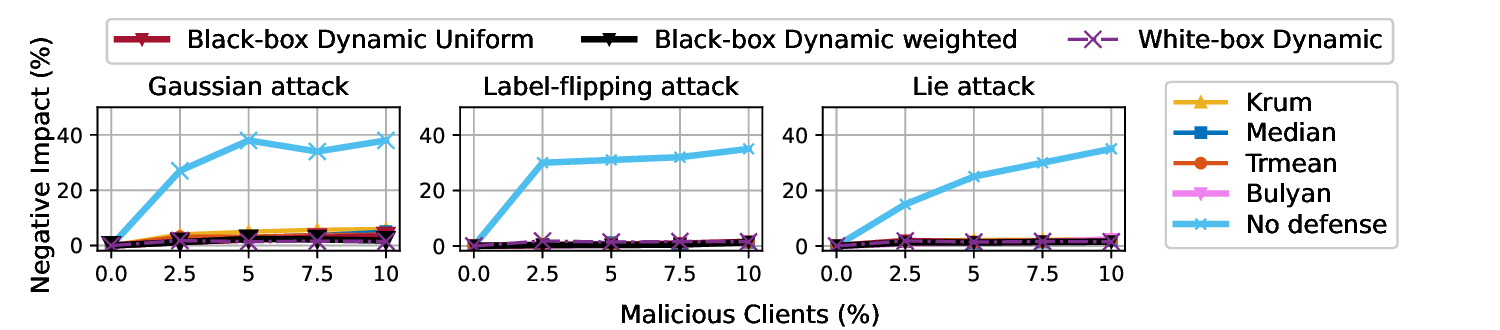}
    \caption{Defending against \texttt{AGR-agnostic} attacks on CIFAR-10. \label{fig:general_attack_cifar10}}
\end{figure}

Figure~\ref{fig:general_attack_cifar10} depicts the attack impacts of three \texttt{AGR-agnostic} attacks to different aggregation algorithms on CIFAR-10 dataset. Similar to the results in Figure~\ref{fig:general_attack_femnist}, as the proportion of malicious clients increases, the negative impacts brought by Byzantine attacks on the models learned without defense become larger, while the performance of models learned with Krum, Median, Trimmedmean, Bulyan, and DDS stays at the same level.  

\end{document}